\documentclass[preprint]{jasatex}
\usepackage{graphicx}
\usepackage{dcolumn}
\usepackage{amsmath,amsfonts}
\usepackage{bm}

\def\bra#1{\mathinner{\langle{#1}|}}
\def\ket#1{\mathinner{|{#1}\rangle}}
\def\braket#1{\mathinner{\langle{#1}\rangle}}
\def\Bra#1{\left\langle#1\right|}
\def\Ket#1{\left|#1\right\rangle}
\let\protect\relax
{\catcode`\|=\active
  \xdef\Braket{\protect\expandafter\noexpand\csname Braket \endcsname}
  \expandafter\gdef\csname Braket \endcsname#1{\begingroup
     \ifx\SavedDoubleVert\relax
       \let\SavedDoubleVert\|\let\|\BraDoubleVert
     \fi
     \mathcode`\|32768\let|\BraVert
     \left\langle{#1}\right\rangle\endgroup}
}
\def\BraVert{\@ifnextchar|{\|\@gobble}
     {\egroup\,\mid@vertical\,\bgroup}}
\def\BraDoubleVert{\egroup\,\mid@dblvertical\,\bgroup}
\let\SavedDoubleVert\relax

{\catcode`\|=\active
  \xdef\set{\protect\expandafter\noexpand\csname set \endcsname}
  \expandafter\gdef\csname set \endcsname#1{\mathinner
        {\lbrace\,{\mathcode`\|32768\let|\midvert #1}\,\rbrace}}
  \xdef\Set{\protect\expandafter\noexpand\csname Set \endcsname}
  \expandafter\gdef\csname Set \endcsname#1{\left\{%
     \ifx\SavedDoubleVert\relax \let\SavedDoubleVert\|\fi
     \:{\let\|\SetDoubleVert
     \mathcode`\|32768\let|\SetVert
     #1}\:\right\}}
}
\def\midvert{\egroup\mid\bgroup}
\def\SetVert{\@ifnextchar|{\|\@gobble}
    {\egroup\;\mid@vertical\;\bgroup}}
\def\SetDoubleVert{\egroup\;\mid@dblvertical\;\bgroup}

\begingroup
 \edef\@tempa{\meaning\middle}
 \edef\@tempb{\string\middle}
\expandafter \endgroup \ifx\@tempa\@tempb
 \def\mid@vertical{\middle|}
 \def\mid@dblvertical{\middle\SavedDoubleVert}
\else
 \def\mid@vertical{\mskip1mu\vrule\mskip1mu}
 \def\mid@dblvertical{\mskip1mu\vrule\mskip2.5mu\vrule\mskip1mu}
\fi

\begin{document}

\title[Diffuse fields in  layered  media]{Normal modes of layered elastic media and application to diffuse fields}

\author{Ludovic Margerin}
\email[e-mail:]{margerin@cerege.fr}
\affiliation{   Centre Europ\'een de Recherche et d'Enseignement  des G\'eosciences de l'Environnement, \\ Universit\'e Aix Marseille, Centre  National de la Recherche Scientifique, \\ Aix en Provence,  France     }

\date{\today}

\begin{abstract}
The spectral decomposition of the elastic wave operator in a layered
isotropic half-space is derived by means of standard functional analytic methods. 
Particular attention is paid to the coupled $P$-$SV$ waves.
The problem is formulated directly in terms of displacements which leads to a $2 \times 2$ Sturm-Liouville 
 system.  The resolvent kernel (Green function)  is expressed in terms of simple plane-wave solutions.
Application of Stone's formula leads naturally
to eigenfunction expansions in terms of generalized eigenvectors with oscillatory behavior at infinity. 
The generalized normal mode expansion is employed to define a diffuse field as a white noise process in modal space.
By means of a  Wigner transform, we calculate vertical to horizontal kinetic energy ratios in layered media,
 as a function of depth and frequency. 
Several illustrative examples are considered including energy ratios near a free surface, in the presence of 
a soft layer. Numerical comparisons between the generalized 
eigenfunction summation and  a classical locked-mode approximation demonstrate the validity  of the approach.  
The impact of the local velocity structure on the energy partitioning of a diffuse field is illustrated.
\end{abstract}

\pacs{43.40.Hb,43.20.Bi,43.40.Ph}

\keywords{  layered media, normal modes, eigenfunction expansion, spectral theory, equipartition, diffuse field}

\maketitle

\section{Introduction}  
Since its introduction in elastodynamics by Weaver \cite{weaver82}, the diffuse
field concept and the principle of equipartition of elastic waves have been successfully applied to both field \cite{hennino01}
and laboratory experiments \cite{malcolm04}. It is an important ingredient of the  spectacular reconstruction
of Green's function from thermal noise \cite{weaver01} or seismic coda waves \cite{campillo03}.
The principle of equipartition can also be used to predict the partition of energy  in diffuse fields
  \cite{weaver82,weaver85,hennino01}. Yet, the practical implementation  of the equipartition principle in a
configuration as simple as a half-space poses some serious technical difficulties \cite{weaver85}.
The simplest mathematical formulation of equipartition relies on the the existence of
a complete normal mode basis. In the past, simple waveguides or a homogeneous half-space
 have been considered \cite{weaver85,tregoures02}.
 In this work, we construct a complete set of normal modes of the elastic wave equation
  from the  general spectral theory of self-adjoint operators in Hilbert space \cite{reed79,pearson88}.
 A diffuse field is then represented as a white noise distributed over the complete set of normal modes independent
of the nature of the spectrum (discrete/continuous), or medium (open/closed).
The paper is organized as follows: in section \ref{problem},  the spectral problem is
introduced and  general properties of the elastodynamic operator are summarized.
In section \ref{stone}, the key mathematical results are presented and illustrated
on a simple problem. Sections \ref{generalized}-\ref{application} present the central
results of the paper. The spectral theory of a layered elastic half-space is derived
in detail and  applied to energy partitioning of diffuse fields.
Other  applications of the theory are briefly discussed in conclusion.

\section{ Problem Statement} \label{problem}
The elastodynamic equation governing seismic wave propagation can be written as
\begin{equation}
\label{elasto1}
 \partial_t^2  \ket{u}   = -\mathbf{L}  \ket{u} ,
\end{equation}
or more explicitly in the the position representation:
 \begin{equation}
\label{elasto2}
 \partial^2_t u_i(\mathbf{x}) = 
        \frac{1}{\rho(\mathbf{x})} \partial_j C_{ijkl} \partial_{k} u_l(\mathbf{x}).
\end{equation}
In equation (\ref{elasto1}), the minus sign has been introduced in order to deal with
 a positive definite operator.

 In an isotropic  medium where properties depend only on the depth
 coordinate $z$, the elastodynamic equation decouples into independent scalar ($SH$) and  vectorial ($P$-$SV$) 
  equations. Our goal is to obtain a complete set of normal modes for the latter problem. Following Ref.\onlinecite{kennett83},
we  reduce the three-dimensional problem to a coupled system of second order differential equations.
 Translational invariance  suggests to look for solutions of the form
 $\mathbf{u}(z) e^{ik x} $. Additionally, cylindrical symmetry enables one to  work in a single
 plane of propagation and ignore the third space dimension. Taking all the symmetries into account yields
  an eigenvalue problem for a second-order  differential operator $\mathbf{L}$ \cite{kennett83}
\begin{equation}~
\label{elasto3}
\left\{
 \begin{aligned}   
 \braket{z | \mathbf{L} |u}  & =   -\frac{1}{\rho(z)}\partial_z \braket{z | \tau | u} + k  B^t \partial_z \braket{z | u} + 
           k^2 C  \braket{z | u}  = \lambda  \braket{z | u} \\  
     \braket{z | \tau | u} & =  A\partial_z \braket{z | u} + k B  \braket{z | u}           
\end{aligned} \right.    .
  \end{equation}  
 The operator $\tau$ provides the tractions generated by the displacement field $\ket{u} $
 and $\rho$ denotes the density.
The matrices $A$, $B$ and $C$ are defined as
\begin{equation}
\begin{split}
  A  = \rho(z) \begin{pmatrix} 
          \alpha^2(z) & 0 \\
          0    & \beta^2(z)    
 \end{pmatrix},   &
\quad
  B  = \rho(z) \begin{pmatrix} 
          0 &   2\beta^2(z) - \alpha^2(z)  \\
         \beta^2(z)   &  0    
 \end{pmatrix},     \\  
 \quad
  C =  \begin{pmatrix} 
          \beta^2(z) & 0 \\
          0    & \alpha^2(z)     &    
 \end{pmatrix} ,
\end{split}   
\end{equation}
where $\alpha$ and $\beta$ denote the longitudinal and shear wavespeed, respectively.
 The  vector  $\ket{u}$ has components $(iu_z,u_x)$, where $u_x$ and $u_z$ denote the
original displacement field. This change of variables can be represented by  a unitary transformation $U$ 
and  turns the original problem into an equivalent, easier problem. Indeed, the multiplication
of the vertical component by $i$ makes $\mathbf{L}$ a real operator
 which simplifies the calculations in the layered case. Once the eigenvectors of the transformed operator have been obtained,
those pertaining to the original operator are recovered by application of the inverse operation $U^{\dagger}$.
As shown previously  \cite{dermenjian85,dermenjian88}, the elastodynamic operator is  positive definite and self-adjoint in the Hilbert
space with    scalar product
\begin{equation}
 \label{scalprod2d}
   \braket{u | v} = \int\!\! dz \rho(z) u_i(z)^* v_i(z)  . 
\end{equation}
The precise domain of definition of $\mathbf{L}$ is given in Ref.~\onlinecite{dermenjian85} but loosely
speaking, the operator acts on functions of finite elastic deformation energy.
In equation (\ref{elasto3}) and (\ref{scalprod2d}) we have introduced the Dirac bra-ket notation for its compactness
and convenience. Since it will be used throughout the paper, we have summarized the main properties of the Dirac formalism in
 appendix \ref{dirac}. 

\section{Stone formula and a simple application} \label{stone}
\subsection{Mathematical results}
The Stone formula \cite{reed80} is a powerful functional analytic result for self-adjoint operators in a Hilbert space. This result
is particularly useful for the  mathematical formulation of scattering  theory 
\cite{reed79,wilcox84,pearson88}. In this paper, we will
formally apply this formula to the elastic wave equation to obtain generalized eigenfunction expansions.
The  eigenvectors $\Ket{e}$  are solutions of the equation $\mathbf{ L} \Ket{e} = \lambda \Ket{e}$
but do not belong to the domain of the operator, i.e., they are not solutions with finite energy.
These eigenvectors are required to construct the complete modal solutions of the elastic wave equation
\cite{maupin96}. Our work  generalizes previous results obtained 
in the case of a homogeneous half-space \cite{secher98,dermenjian85}.  
Before stating the Stone formula, we introduce the resolvent $\mathbf{G}(\lambda)$ of the operator $\mathbf{L}$:
 \begin{equation}
   \mathbf{G} = \left( \mathbf{L} - \lambda \mathbf{I} \right)^{-1}   .
 \end{equation}
Because $\mathbf{L}$ is self-adjoint, its spectrum is a subset of the real axis 
and therefore the resolvent is  defined for $ \text{Im}\lambda \neq 0$.
 In position and polarisation space, the resolvent is represented by the 
 Green matrix 
 \begin{equation}
   \braket{ z | \left( \mathbf{L} - \lambda \mathbf{I} \right)^{-1} | z' }=
      G_{ij}\left(\lambda, z,z' \right),
 \end{equation}
which possesses the Hermitian symmetry
 \begin{equation}
     \label{symmetry}
      G_{ij}\left(\lambda, z,z' \right) =  G_{ji}\left(\lambda^*, z',z \right)^* .
 \end{equation}
We first recall the fundamental spectral theorem and the Stone formula in an abstract setting. 
 To every self-adjoint operator $\mathbf{L}$ one can associate a projection operator valued
 function (measure) $\mathbf{P}_{\lambda}$ having the following properties \cite{reed80}
 \begin{equation}
   \mathbf{f(L)} = \int\limits_{-\infty}^{+\infty} f(\lambda) d\mathbf{P}_{\lambda} ,
 \end{equation} 
The spectral theorem guarantees that the spectral family $\mathbf{P}_{\lambda}$ is orthogonal,
 diagonalizes the operator $\mathbf{L}$ and provides the completeness relation.
 A practical method to construct the spectral projector  $\mathbf{P}_{\lambda}$
 is provided by the Stone formula, which relates functions of a self-adjoint operator to the discontinuity of its resolvent across
 the real axis \cite{reed80}:
 \begin{equation}
 \label{stone1}
     f\left(\mathbf{L} \right) = \frac{1}{2\pi i} \lim\limits_{\epsilon \to 0^+} \int\limits_{-\infty}^{+\infty}
       \left[\mathbf{G}\left(\lambda +i\epsilon \right) -  \mathbf{G}\left(\lambda - i\epsilon \right)   \right] f(\lambda) d\lambda  .
 \end{equation}
For a positive operator, the integral can be taken from $0$ to $+\infty$ only because the spectrum is a subset
of the positive real axis. As a particular case of equation (\ref{stone1}), one can formally obtain the completeness relation:
\begin{equation}
   \label{resid}
     \mathbf{I}   = \frac{1}{2\pi i} \lim\limits_{\epsilon \to 0^+} \int\limits_{-\infty}^{+\infty}
       \left[\mathbf{G}\left(\lambda +i\epsilon \right) -  \mathbf{G}\left(\lambda - i\epsilon \right)   \right]  d\lambda  .
 \end{equation}
For an operator with a mixed spectrum, with both discrete and  continuous parts,  equation
  (\ref{resid}) will  take  the following form:
\begin{equation}
    \label{resid2}
   \mathbf{I} = \int d\lambda \sum\limits_{m} \Ket{e_{m}(\lambda)} \Bra{e_{m}(\lambda)} 
                            +  \sum\limits_n   \Ket{e_{n}} \Bra{e_{n}}  ,
\end{equation} 
   where $ \ket{e_{m}(\lambda)} $ denotes generalized eigenfunctions of the continuum -we assume that no singular continuous spectrum
 exists- that are normalized in the following sense:
  \begin{equation}
    \braket{e_{m}(\lambda) | e_{m'}(\lambda')}  = \delta_{m m'} \delta(\lambda - \lambda').
  \end{equation}
In equation (\ref{resid2}), we have introduced a discrete index in the continuous part of the spectrum, which corresponds to the
possibility that the space of generalized eigenfunction be multidimensional.   
\subsection{Generalized eigenvectors in homogeneous full-space}   
We consider the system (\ref{elasto3})  with constant parameters $\alpha_{\infty}, \beta_{\infty}$ and $\rho_{\infty}$. First we construct the
resolvent $\mathbf{L} -\lambda \mathbf{I}  $ outside of the real axis. In order to uniquely define the square root 
of a complex number $z$ we adopt the following convention: 
  $\sqrt{z} = \lvert z\rvert^{1/2} e^{i \theta/2} ,$ with $ \theta = \arg(z) \in \left[0, 2 \pi \right).$ This definition guarantees
 that $\text{Im}\sqrt{z} > 0 $ when $z$ lies outside the positive real axis. The $2\times2$ system of coupled second-order
equation (\ref{elasto3}) has four linearly independent solutions $\Ket{ u^P_{+\infty}(\lambda)} $,
 $\Ket{ u^P_{-\infty}(\lambda) }$,
$\Ket{ u^S_{+\infty}(\lambda) }$, $\Ket{ u^S_{-\infty}(\lambda) }$, whose properties are summarized below:
\begin{equation}
 \label{basis}
 \begin{split}
    \braket{ z |  u^P_{\pm\infty}(\lambda)  } & = 
      \frac{1}{\epsilon_{\alpha}(\lambda)} \left( 
    \begin{gathered}
     \pm i q_{\alpha}(\lambda)  \\
       k  
      \end{gathered}  \right) e^{\pm i q_{\alpha}(\lambda) z} ,   
    \\
   \braket{ z |  u^S_{\pm\infty}(\lambda)  }  & = 
      \frac{1}{\epsilon_{\beta}(\lambda)} \left( 
    \begin{gathered}
            k      \\
     \mp i q_{\beta}(\lambda)    
      \end{gathered}  \right) e^{\pm i q_{\beta}(\lambda)z}    .
\end{split}
\end{equation}
In equation (\ref{basis}), the prefactors $\epsilon_{\alpha,\beta}$  ensure that the plane waves are energy normalized \cite{kennett83}:
 \begin{equation}
    \epsilon_{\alpha,\beta}  = \frac{1}{\sqrt{2 \rho_{\infty}\sqrt{\lambda} q_{\alpha,\beta}} },
\end{equation}
 and the vertical  wavenumbers $q_{\alpha,\beta}$ are defined as 
 \begin{equation}
   q_{\alpha}  = \sqrt{\frac{\lambda}{\alpha_{\infty}^2} -k^2} \text{, \quad }  
 q_{\beta}  =  \sqrt{\frac{\lambda}{\beta_{\infty}^2} -k^2} .
\end{equation}     
The following symmetry relations will prove useful:
\begin{equation}
  \label{symmetry2}
  \begin{split}
    \braket{ z |  u^P_{+\infty}(\lambda^*)  } & = -  \braket{ z |  u^P_{+\infty}(\lambda)  }^* \\
   \braket{ z |  u^P_{-\infty}(\lambda^*)  } & = -  \braket{ z \vert  u^P_{-\infty}(\lambda)  }^*
 \end{split},
\end{equation}
with completely analogous relations involving $\Ket{ u^S_{+\infty}(\lambda) }$, $\Ket{ u^S_{-\infty}(\lambda) }$.
When $\text{Im}\lambda=0 $, $\lambda>\alpha_{\infty}^2 k^2$, assuming a time dependence of the form $e^{-i \sqrt{\lambda} t}$,
 $( u^P_{+\infty} , u^P_{-\infty} )$--$( u^S_{+\infty},  u^S_{-\infty}  )$ behave like (outgoing,incoming) plane $P$--$S$ waves
 at $+\infty$, respectively. 
Using this set of solutions, we can construct the Green matrix for $\text{Im} \lambda \neq 0$, which obeys the following
 equation:
 \begin{equation}
  \label{diffeqg}
    -\rho^{-1}_{\infty} \partial_z\tau[G]_{ij}(z,z') + K_{ik}(z) G_{kj}(\lambda,z,z') 
      - \lambda  G_{ij}(z,z')    = \rho^{-1}_{\infty} \delta_{ij} \delta(z-z')    ,
 \end{equation}
where we have introduced the differential operator $K(z) =  k^2 C + k B^t \partial_z   $.
The Green matrix should possess the following properties:
 \begin{enumerate}
  \item  For $z>z'$ , each column of $G$  must be a linear combination of   solutions with finite
      energy at $ +\infty$ $(\ket{u^P_{+\infty}},\ket{u^S_{+\infty})}$.
  \item  For $z<z'$ , each column of $G$  must be a linear combination of    solutions
        with finite energy    at $-\infty$  $(\ket{u^P_{-\infty}},\ket{u^S_{-\infty}})$.  
    \item $G$ obeys the symmetry relation (\ref{symmetry})  
  \item  $G$ is continuous on the diagonal  $z=z'$.
  \item The traction matrix of $G$ has a jump discontinuity on the diagonal:
             $\tau[G]_{ij}{\displaystyle \arrowvert^{z^{'+}}_{z^{'-}}}  =  -\delta_{ij}$  
\end{enumerate}
Properties (1)-(5) are straightforward generalizations to systems of 2$^{nd}$ order differential
 equations of the standard Sturm-Liouville theory \cite{secher98}. Properties (1)-(3) indicate that
 $\mathbf{G}(\lambda)$ should have  the following form
\begin{equation}
\label{ginf}
 \mathbf{G(\lambda)} = \begin{cases}  
   &   K_1 (\lambda) \Ket{  u^P_{-\infty}(\lambda) }  \Bra{ u^P_{+\infty}(\lambda)^* }  +
       K_2 (\lambda) \Ket{  u^S_{-\infty}(\lambda) }  \Bra{ u^S_{+\infty}(\lambda)^* }   \quad z< z'  \\
  &   K_1 (\lambda) \Ket{  u^P_{+\infty}(\lambda)  } \Bra{ u^P_{-\infty}(\lambda)^*  }   +   
       K_2 (\lambda) \Ket{  u^S_{+\infty}(\lambda) } \Bra{ u^S_{-\infty}(\lambda)^*  }       \quad z> z'
\end{cases}.
\end{equation}
In equation (\ref{ginf}) we have separated the Green matrix into a $P$ and an $S$ wave part.
Application of the continuity condition of the Green matrix on the diagonal (4)  yields
$K_1(\lambda) = K_2(\lambda)$.
Next we apply the  condition (5) for the discontinuity of the traction on the diagonal  
to obtain: 
\begin{equation}
  K_1(\lambda) = K_2(\lambda) = \frac{i}{\sqrt{\lambda}} .
\end{equation}


The next step is to study the discontinuity of the resolvent across the real axis. 
 We will denote by
$\ket{ u_{+\infty}^{P+} }$ and $\ket{ u_{+\infty}^{P-} }$ the limiting values of the vectors
$\ket{ u_{+\infty}^{P}(\lambda) }$ as $\lambda$ approaches the real axis from above and below, respectively,
with analogous definitions for $\ket{ u_{-\infty}^{P+} }$ and $\ket{ u_{+\infty}^{S+}}$, $\cdots$.
The jump of the resolvent $\mathbf{G^+} - \mathbf{G^-}$ across the real axis depends on the relations among  these 
functions.
For $z<z'$, one obtains 
\begin{equation}
\label{disc1}
  \mathbf{G^+} - \mathbf{G^-} = \frac{i}{\sqrt{\lambda^+}} 
               \left( \ket{ u_{-\infty}^{P+} } \bra{ ( u_{+\infty}^{P+})^*  } +  
                \ket{ u_{-\infty}^{P-} } \bra{ ( u_{+\infty}^{P-})^*  }    + 
                  \ket{ u_{-\infty}^{S+} } \bra{ ( u_{+\infty}^{S+})^* }   +  
                \ket{ u_{-\infty}^{S-} } \bra {( u_{+\infty}^{S-})^*  }  \right).
\end{equation}
Three cases have to be distinguished: \\
(1) $\lambda > \alpha_{\infty}^2 k^2 $.  In this case $q_{\alpha,\beta}$ and $\epsilon_{\alpha,\beta}$ are real and all the waves
 are propagating.  Applying the following symmetry relations,
\begin{equation}
\begin{aligned}
\ket{ ( u_{+\infty}^{P+})^*  } = & \ket{ u_{-\infty}^{P+} } \quad  & 
 \ket{  u_{+\infty}^{P-}  } = & - \ket{ (u_{+\infty}^{P+})^* }  \\
\ket{ ( u_{+\infty}^{S+})^*  } = & \ket{ u_{-\infty}^{S+} } \quad  &
\ket{  u_{+\infty}^{S-}  } = & - \ket{ (u_{+\infty}^{S+})^* },
\end{aligned}
\end{equation}
one can rewrite the resolvent discontinuity (\ref{disc1}) as a sum of projection operators on a
 generalized eigenfunction basis:
 \begin{equation}
\label{disc2}
  \mathbf{G^+} - \mathbf{G^-} = \frac{i}{\sqrt{\lambda^+}} 
            \left(
                \ket{ u_{-\infty}^{P+} } \bra{ u_{-\infty}^{P+} }  +  
                \ket{ u_{+\infty}^{P+} } \bra{ u_{+\infty}^{P+} }  + 
                 \ket{ u_{-\infty}^{S+} } \bra{ u_{-\infty}^{S+}}  +  
                \ket{ u_{+\infty}^{S+} } \bra{  u_{+\infty}^{S+}}  \right) .
\end{equation}
From equation (\ref{disc2}), one concludes that there are four linearly independent generalized eigenfunctions
 with eigenvalue $\lambda > \alpha_{\infty}^2 k^2$.  For instance, the first term on the left-hand side of equation
(\ref{disc2}) is an orthogonal projector on the space of upgoing $P$ waves. \\
(2 ) $\alpha_{\infty}^2 k^2 > \lambda > \beta_{\infty}^2 k^2 $.  $q_\alpha$ is pure imaginary and $q_{\beta}$ is real which corresponds
to evanescent $P$ waves and propagating $S$ waves. Upon application of the following symmetry relations:
\begin{equation}
\begin{aligned}
\ket{ ( u_{-\infty}^{P-})  } & =  -i\ket{ u_{-\infty}^{P+} }  \quad &   
 \ket{  u_{+\infty}^{P-}  }  & =  -i \ket{ u_{+\infty}^{P+} }   \\
\ket{ ( u_{+\infty}^{S+})^*  } & =  \ket{ u_{-\infty}^{S+} }  \quad  &
\ket{  u_{+\infty}^{S-}  }  & =  - \ket{ (u_{+\infty}^{S+})^* } ,
\end{aligned}
\end{equation} 
equation (\ref{disc1}) simplifies to
\begin{equation}
\label{disc3}
  \mathbf{G^+} - \mathbf{G^-} = \frac{i}{\sqrt{\lambda^+}} 
            \left(
            \ket{ u_{-\infty}^{S+} } \bra{ u_{-\infty}^{S+}  }  +  
            \ket{ u_{+\infty}^{S+} } \bra{ u_{+\infty}^{S+}  } \right) .
\end{equation}
There are two linearly independent $S$ eigenfunctions. \\
(3) $\lambda < \beta_{\infty}^2 k^2$. There are no singularities and the resolvent is continuous:
  \begin{equation}
\label{disc4}
  \mathbf{G^+} - \mathbf{G^-} =  \mathbf{0}. 
\end{equation}
Because of the symmetry of the Green's function (\ref{symmetry}), the same results are obtained for $z>z'$.
The completeness relation now follows from the Stone formula (\ref{resid}). The continuous parameter $\lambda$ is
 analogous to a squared eigenfrequency. Introducing $\lambda=\omega^2$, and generalized
eigenvectors $\Ket{ e^P_{+\infty}(\omega) }$, $\cdots$ the resolution of the identity writes as a sum of
integrals over frequencies:
\begin{equation}
\label{resom}
\begin{split}
\mathbf{I} = & \int\limits^{+\infty}_{\alpha_{\infty} k} d\omega  
    \left( \Ket{ e^P_{+\infty}(\omega) } \Bra{ e^P_{+\infty}(\omega) }  +
         \Ket{ e^P_{-\infty}(\omega) } \Bra{ e^P_{-\infty}(\omega) } \right) + \\
       & \int\limits^{+\infty}_{\beta_{\infty} k} d\omega  
    \left( \Ket{ e^S_{+\infty}(\omega) } \Bra{ e^S_{+\infty}(\omega) } +
         \Ket{ e^S_{-\infty}(\omega) } \Bra{ e^S_{-\infty}(\omega) } \right) .
\end{split}
\end{equation}
The eigenvectors $\Ket{ e^P_{+\infty}(\omega) }$, $\cdots$ can be deduced from the basis vectors
  $\Ket{ u^P_{+\infty}(\lambda) }$ and form an orthonormal set. Let us for instance consider
the eigenvector $ \ket{e^P_{+\infty}(\omega) }$:
 \begin{equation}
 \braket{ z  | e^P_{+\infty}(\omega) } = \frac{1}{\sqrt{2 \pi \rho_{\infty} \omega q_{\alpha}}} 
                      \left(
 \begin{gathered}   -i q_\alpha  \\
                      k
 \end{gathered} 
          \right) e^{+iq_{\alpha}z},   \quad\quad   \braket{  e^P_{+\infty}(\omega)  | e^P_{+\infty}(\omega') } =
          \delta(\omega - \omega'),
 \end{equation}
where $q_{\alpha} = \sqrt{\frac{\omega^2}{\alpha^2_{\infty}} -k^2}$. 
Multiplying the wavefunctions $\braket{ z  | e^P_{+\infty}(\omega) } $ by the phase term $\frac{1}{\sqrt{2 \pi}}e^{ikx}$ 
we can  form compound eigenvectors $\ket{ \psi^{P,S} }$ that can be used to expand 2-D in-plane vector fields. 
The expansion takes a particularly simple form if instead of using a frequency integral, the new variables
$p_z = \sqrt{\frac{\omega^2}{\alpha^2_{\infty}} -k^2}$ and $p_z = \sqrt{\frac{\omega^2}{\beta^2_{\infty}} -k^2}$
are introduced  in the $P$ and $S$ integrals  of equation (\ref{resom}), respectively. In order to obtain
eigenfunctions of the elastodynamic operator in its standard form, we apply the unitary transformation
 $U^{\dagger}$ to obtain a fairly simple completeness relation:
\begin{equation}
    \mathbf{I} =  \iint\limits_{\mathbb{R}^2} dp_xdp_z \left( \Ket{ \psi^P(p_x,p_z) } \Bra{ \psi^P(p_x,p_z)  }  
              +  \Ket{ \psi^S(p_x,p_z) } \Bra{ \psi^S(p_x,p_z) }   \right) ,
\end{equation} 
 where $\Ket{ \psi^P(p_x,p_z) }$ and $\Ket{ \psi^S(p_x,p_z) }$ are simple plane $P$ and $SV$ waves:
\begin{equation}
\label{psips}
\begin{split}
 \braket{ r | \psi^P(p_x,p_z) } = & \hat{p}\frac{e^{i p_x x + i p_z z}}{2 \pi \sqrt{\rho_{\infty}}} \\
   \braket{ r | \psi^S(p_x,p_z) } = & \hat{p}^{\perp}\frac{e^{i p_x x + i p_z z}}{2 \pi \sqrt{\rho_{\infty}}},
\end{split}
\end{equation}
where $ \hat{p} $ and  $ \hat{p}^{\perp} $ denote unit vectors parallel and perpendicular to the 
wavevector $\mathbf{p}$, respectively:
 \begin{equation}
\label{hatps}
  \hat{p} = \frac{1}{\sqrt{p_x^2 + p_z^2}}\left( \begin{gathered} p_x \\
                                             p_z
\end{gathered}
  \right), \quad
  \hat{p}^{\perp} = \frac{1}{\sqrt{p_x^2 + p_z^2}}\left( \begin{gathered} -p_z \\
                                             p_x
\end{gathered}
  \right).
 \end{equation}
 $\Ket{ \psi^P(p_x,p_z) }$ and $\Ket{ \psi^S(p_x,p_z) }$  are eigenvectors of the elastic  wave equation
with eigenvalues $\alpha^2_{\infty} (p_x^2 + p_y^2)  $ and $\beta^2_{\infty} (p_x^2 + p_y^2) $, respectively. They
form a complete set of normal modes that decompose 2-D vector fields into  longitudinal and transverse parts.
\section{Generalized eigenfunctions in stratified media} \label{generalized}
\subsection{Construction of the resolvent}
We begin with the construction of the resolvent, which  parallels the development of the previous section.
To make comparison easier we denote by $\rho_{\infty}$, $\alpha_{\infty}$ and $\beta_{\infty}$ the
density and seismic wave speeds in the half-space located at depth $z>0$. The medium is assumed
to be composed of a stack of $n$ layers and is bounded by a free surface at $z=-z_n$, $z_n>0$.
We wish to find the eigenfunction expansion associated with 
the  eigenvalue problem $\mathbf{L}\Ket{u} = \lambda \Ket{u}$ 
introduced in equation (\ref{elasto3}) and  supplemented with the free surface traction
condition $\braket{ z | \tau |u} =0 $ at $z=0$ and the continuity of stresses and displacements at each interface. 
To do so, we introduce four linearly independent vector solutions of the elastic wave equation (\ref{elasto3}) denoted by 
  $\Ket{ u^P_{+\infty}(\lambda) }$, $\Ket{ u^P_{-\infty}(\lambda) }$,
$\Ket{ u^S_{+\infty}(\lambda) }$, $\Ket{ u^S_{-\infty}(\lambda) }$. In the half-space $z>0$, their analytical
form is given by  equation (\ref{basis}). Their analytical dependence in the stack of layer is obtained by integrating
the equations of motion (\ref{elasto3}) from depth $z=0^{+}$ to $z=-z_n^{+}$ by applying continuity of stresses and
displacements at each interface. Note that, in general, these solutions do \emph{not} verify the stress free condition at $z=0$. 
Following Ref.\onlinecite{kennett83}, we introduce two linearly independent solutions  $\Ket{u_0^P(\lambda)}$,
$\Ket{u_0^S(\lambda)} $ of the wave equation that verify the stress-free condition at $z=0$
\begin{equation}
\label{u0ps}
\begin{split}
 \Ket{u_0^P(\lambda)} = &  \Ket{u^P_{-\infty}(\lambda) } + r^{pp}(\lambda) \Ket{u^P_{+\infty}(\lambda) }
                         + r^{ps}(\lambda) \Ket{u^S_{+\infty}(\lambda) } \\
  \Ket{u_0^P(\lambda)} = &  \Ket{u^S_{-\infty}(\lambda) } + r^{sp}(\lambda) \Ket{u^P_{+\infty}(\lambda) }
                         + r^{ss}(\lambda) \Ket{u^S_{+\infty}(\lambda) } 
\end{split},
\end{equation}
where $r^{pp}$, $r^{ps}$, $r^{sp}$ and $r^{ss}$ are the reflection coefficients of the stack of layer including
the free surface. They are uniquely specified by the boundary conditions and the 
form of the solutions in the half-space.
For $\text{Im}\lambda =0 $, $\text{Re} \lambda > \alpha^2_{\infty}k^2 $, the solutions (\ref{u0ps}) behave like propagating $P$ and $S$ waves incident from $+\infty$
together with their reflections. They are solutions of the wave equation with infinite energy that verify
the free surface condition. 
For $\text{Im}\lambda \neq 0$, we remark that because the matrix elements of $\mathbf{L}$ are real, the following
 relation holds:
 $\mathbf{L}\Ket{u_0^P(\lambda)^*} = \lambda^* \Ket{u_0^P(\lambda)}  = \mathbf{L}\Ket{u_0^P(\lambda^*)} $.
Using equation (\ref{symmetry2}), which is still valid in the stratified case, 
the following symmetry relations are established:
\begin{equation}
\ket{u^{P,S}_0(\lambda^*)}  =  -\ket{u^{P,S}_0(\lambda)}, 
\end{equation}
\begin{equation}
\label{symsemi1}
   r^{ab}(\lambda^*) =  r^{ab}(\lambda)^* , 
\end{equation}
where $a,b=p,s$.  In addition, using the  method of propagation invariants developed in Ref.\onlinecite{kennett78},
 we can prove the following reciprocity relation:
       \begin{equation}
  \label{reciprocity}
    r^{ps}(\lambda) = r^{sp}(\lambda).
       \end{equation}
 To construct the   Green matrix we  again make use of  the properties (1)-(3) above. Condition (1) is replaced by the
 requirement that the columns of $G$ be linear combinations of the vectors $\ket{u_0^P}$, $\ket{u_0^S}$ for $z<z'$.
This suggests to try the following  form for $\mathbf{G}(\lambda)$:
\begin{equation}
 \label{gsemi}
 \mathbf{G(\lambda)} =  \begin{cases}  
   & K_1 \Ket{  u^P_0(\lambda)  }  \Bra{ u^P_{+\infty}(\lambda)^* }
      + K_2 \Ket{  u^S_0(\lambda)  }  \Bra{ u^S_{+\infty}(\lambda)^* }  \quad z<z'  \\
  &  K_1  \Ket{  u^P_{+\infty}(\lambda) } \Bra{ u^P_0(\lambda)^*  }    
      +  K_2  \Ket{  u^S_{+\infty}(\lambda)  }  \Bra{ u^S_0(\lambda)^* }     \quad z>z'
\end{cases}    ,
\end{equation}
where use has been made of equation (\ref{symsemi1}). 
In essence, equation (\ref{gsemi}) is  completely similar to equation  (\ref{ginf}): the Green function is separated
into a ``$P$'' and an ``$S$'' wave part, although the vectors $\ket{u_0^P}$, 
$\ket{u_0^{S}}$ are neither purely longitudinal nor
purely transverse, due to the mode conversions at the medium boundaries.
We must now determine the values of $K_1$ and $K_2$ in order to fulfill conditions (4)-(5). When the source
is located in the half-space $z>0$, we can make  use of the analytical form of the vectors $\ket{u_0^{P,S}}$ and
$\ket{u^{P,S}_{+\infty}}$. Following the same steps as in the homogeneous case and employing the reciprocity relation 
(\ref{reciprocity}) one  obtains:
\begin{equation}
 K_1(\lambda) = K_2(\lambda) = \frac{i}{\sqrt{\lambda}} .
\end{equation}
Let us now show that this expression is valid throughout the stack of layers.
The key point is to verify conditions (4)-(5).
Let us first introduce the following 4-dimensional displacement-stress vectors :
 \begin{equation}
  w^{P,S}_{\infty}(\lambda,z) = \left( \begin{gathered}  u^{P,S}_{+\infty}(\lambda,z)     \\
                 \tau[u^{P,S}_{+\infty}(\lambda,z)]    
      \end{gathered}        \right) ,           
\end{equation}
and
\begin{equation}
 w^{P,S}_{0}(\lambda,z) = \left( \begin{gathered}  u^{P,S}_{0}(\lambda,z)     \\
                 \tau[u^{P,S}_{0}(\lambda,z)]    
      \end{gathered}        \right) .
\end{equation}  
 The conditions (4)-(5)  will be satisfied provided the following
relation holds at arbitrary depth $z$ in the medium:
\begin{equation}
  \label{system3}
\begin{aligned}
 \frac{i}{\sqrt{\lambda}}  & \left(  w^{P}_{\infty}(\lambda,z)  w^{P}_{0}(\lambda,z)^t -  w^{P}_{0}(\lambda,z)  w^{P}_{\infty}(\lambda,z)^t \right. \\
         &  \left. \quad  +   w^{S}_{\infty}(\lambda,z)  w^{S}_{0}(\lambda,z)^t -  w^{S}_{0}(\lambda,z)  w^{S}_{\infty}(\lambda,z)^t \right) =
    N ,
\end{aligned}
\end{equation}  
where we have introduced the notations :
\begin{equation}
N=  \begin{pmatrix}
    0 & - E \\
   E &  0
   \end{pmatrix}    \text{ and } \quad      
 E = \begin{pmatrix}
  1 & 0 \\
  0 & 1
 \end{pmatrix} .
\end{equation}
Relation (\ref{system3}) is verified in the half-space $z>0$ as discussed above. 
 The displacement-stress vector satisfies a first-order linear system 
$   d w(\lambda,z)/dz = A(\lambda,z) w(\lambda,z)$, equivalent
 to the second order system (\ref{elasto3}) , 
which can be integrated with the aid of a propagator matrix:
\begin{equation}
   w(\lambda,z) = P(\lambda,z,z') w(\lambda,z').
\end{equation}
Since the propagator matrix has  the symplectic symmetry \cite{kennett83}:
\begin{equation}
\label{symplectic}
   P(\lambda,z,z') N P(\lambda,z,z')^t = N,
\end{equation}
relation (\ref{system3}) can be propagated at any depth in the stratified half-space by left
and right multiplication of equation (\ref{system3}) by $P$ and its transpose, respectively.
Note that in the case of a stratified medium,  the symmetry relation (\ref{symplectic}) can be
verified by direct but tedious calculations.

\subsection{Singularities of the resolvent}
\begin{figure*}
  \setlength{\unitlength}{1cm}
\begin{picture}(14,6)
 \put(0,1){\line(1,0){13}}
  \put(1,0){\vector(0,1){6}}
  \put(5,1){\circle*{0.2}}
  \put(9,1){\circle*{0.2}}
  { \linethickness{0.3mm}   \put(5,1){\line(1,0){7}} }
 {  \linethickness{0.6mm} \put(9,1){\line(1,0){4.9}} }
 \put(13.9,1){\vector(1,0){0.1}}
 \put(1.2,5.8){$\text{Im} \lambda$}
\put(13.7,0.3){$\text{Re} \lambda$}
 \put(4.8,0.5){$\beta_{\infty}^2 k^2$}
\put(8.8,0.5){$\alpha_{\infty}^2 k^2$}
 \put(5.05,3.5){$\overbrace{\text{\hspace*{9cm}}}^{\displaystyle \text{Continuous Spectrum}}$}
 \put(1.0,3.5){$\overbrace{\text{\hspace*{3.95cm}}}^{\displaystyle \text{Discrete Spectrum}}$}
 \put(5.3,3){1 eigenfunction}
 \put(9,3){2 linearly independent}
  \put(9,2.7){eigenfunctions}
 \put(6.3,2.8){\vector(0,-1){1.6}}
  \put(10.3,2.5){\vector(0,-1){1.3}}
  \put(1.8,1){\circle*{0.1}}
  \put(2.8,1){\circle*{0.1}}
  \put(3.55,1){\circle*{0.1}}
  \put(4.05,1){\circle*{0.1}}
  \put(4.55,1){\circle*{0.1}}
 \put(1.8,3){Rayleigh modes}
  \put(2.8,2.8){\vector(0,-1){1.6}}
\end{picture}
\caption{Spectral properties of the elastodynamic operator  in a layered half-space. $\lambda$
    is the eigenparameter. } \label{fig:specprop2}
\end{figure*}
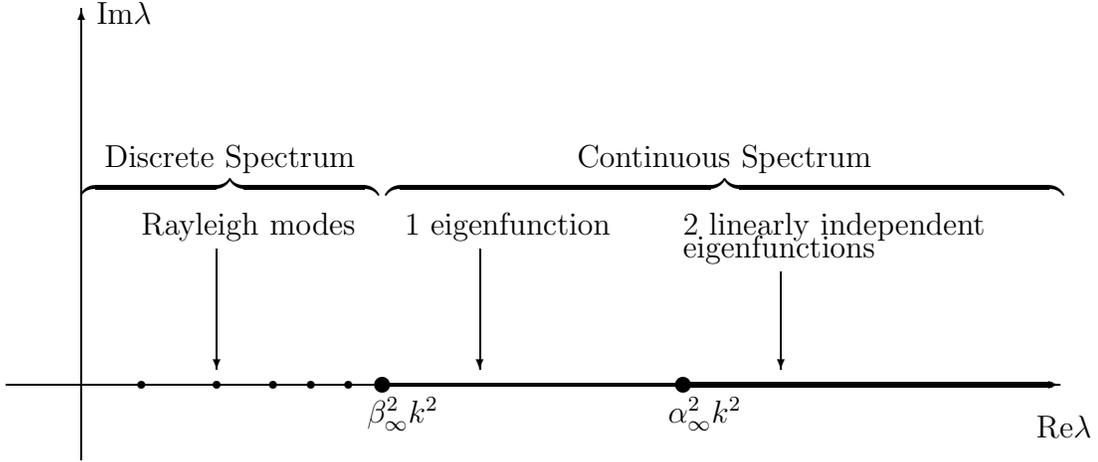

The next step is to reexamine 
the discontinuity of the resolvent (\ref{gsemi}) across the real axis. This task is significantly facilitated by the
unitary and symmetry properties  of the reflection coefficients.
 As shown in Ref.\onlinecite{kennett78}, the energy normalization of the eigenvectors is of fundamental importance in this respect.     
Other normalizations are of course possible but they lead to much more awkward calculations.
We will denote by $\ket{u(\lambda^+)}$-($\ket{u(\lambda^-)}$) the limiting value of the ket
$\ket{u(\lambda)}$ as $\lambda$ approaches the real axis from above-(below). The interrelations
among the kets $\ket{u^{P,S}_{+\infty}}$, $\ket{u^{P,S}_{-\infty}}$ across the real axis determine the form of the resolvent
discontinuity. These relations, in turn, depend essentially on the analytical forms of these kets, i.e., on whether the $P$ and $S$ 
waves are propagating or evanescent in the half-space $z>0$. 
As in the homogenenous case, this leads to a separation of the real axis into  three distinct domains,
as shown in Figure \ref{fig:specprop2}. \\
(1) For propagating $P$ and $S$ waves: $\lambda > \alpha^2 k^2$, the following relations apply:
\begin{equation}
\begin{aligned}
 \Ket{u^{P}_{\pm\infty}(\lambda^+)}  = &   \Ket{u^{P}_{\mp\infty}(\lambda^+)^*} & 
   \Ket{u^{S}_{\pm\infty}(\lambda^+)}  = & \Ket{u^{S}_{\mp\infty}(\lambda^+)^*}  \\
 \Ket{u^{P}_{\pm\infty}(\lambda^-)}  = &   -\Ket{u^{P}_{\pm\infty}(\lambda^+)^*}   & \Ket{u^{S}_{\pm\infty}(\lambda^-)}  = &
       - \Ket{u^{S}_{\pm\infty}(\lambda^+)^*}    
 \end{aligned}   .
\end{equation}    
These relations, together with the analytical properties of the reflection coefficients,
 enable us to write the kets $\ket{u_0^{P,S}(\lambda^{-})}$ in terms of the kets $\ket{u^{P,S}_{\pm\infty}(\lambda^+)}$ only:
\begin{equation}
\begin{aligned}
 \label{defu0}
 \Ket{u_0^P(\lambda^-)} & =  -\Ket{u_{+\infty}^P(\lambda^+)} - \overline{r^{pp}} \Ket{u_{-\infty}^P(\lambda^+)}  
            -  \overline{r^{ps}} \Ket{u_{-\infty}^S(\lambda^+)} , \\
    \Ket{u_0^S(\lambda^-)} & =  -\Ket{u_{+\infty}^S(\lambda^+)} - \overline{r^{ss}} \Ket{u_{-\infty}^S(\lambda^+)}  
            -  \overline{r^{sp}} \Ket{u_{-\infty}^P(\lambda^+)} ,   
\end{aligned}
\end{equation}
where the reflection coefficients are defined by their limiting value as $\lambda$ approaches the real axis from above.
For notational convenience, we have denoted the complex conjugation with an overbar.
When both $P$ and $S$ waves are propagating in the half-space, the matrix of reflection coefficients is unitary  \cite{kennett78}:
\begin{equation}
\label{unitary}
     R = \begin{pmatrix}
     r^{pp} & r^{ps} \\
     r^{sp} & r^{ss}
     \end{pmatrix}      , \quad  R R^{\dagger} = E.
\end{equation} 
For $z<z'$, the discontinuity of the resolvent writes:
\begin{equation}
\begin{split}
      \mathbf{G}^+ -  \mathbf{G}^-           &        =   \frac{i}{\sqrt{\lambda^+}} \left(
                            \Ket{u_0^P(\lambda^+)} \Bra{u_{-\infty}^P(\lambda^+)} + \Ket{u_0^S(\lambda^+)}\Bra{u_{-\infty}^S(\lambda^+)} \right. \\
               & \quad      \left.  -     \Ket{u_0^P(\lambda^-)} \Bra{u_{+\infty}^P(\lambda^+)} - \Ket{u_0^S(\lambda^-)}\Bra{u_{+\infty}^S(\lambda^+)} \right)
\end{split}.
\end{equation}
Using equations (\ref{defu0}) and (\ref{unitary}) and after some algebra, one obtains:
\begin{equation}
\mathbf{G}^+ -  \mathbf{G}^-   =  \frac{i}{\sqrt{\lambda^+}} \left( \Ket{u_0^P(\lambda^+)} \Bra{u_0^P(\lambda^+)}
                           +  \Ket{u_0^S(\lambda^+)} \Bra{u_0^S(\lambda^+)} \right).
\end{equation}
The same result applies for $z>z'$ and we have therefore put the discontinuity of the resolvent in suitable dyadic form.
It appears that for $\lambda > \alpha^2k^2$ there are two linearly independent eigenvectors. They are energy
 normalized incident $P$ and $S$ waves incident from $+\infty$ together with their reflections. \\
(2) For $\alpha^2 k^2 > \lambda > \beta^2 k^2$, there are propagating $S$ and evanescent $P$ waves in the
 half-space $z>0$  and the following relations apply:
\begin{equation}
\label{irel}
\begin{aligned}
 \Ket{u^{P}_{\pm\infty}(\lambda^+)}  = &  -i \Ket{u^{P}_{\pm\infty}(\lambda^+)^*} &  \text{ and } 
   \Ket{u^{S}_{\pm\infty}(\lambda^+)}  = & \Ket{u^{S}_{\mp\infty}(\lambda^+)^*},
  \end{aligned}   
\end{equation}    
which yields:
\begin{equation}
\begin{aligned}
 \label{defu02}
 \Ket{u_0^P(\lambda^-)} & =  -i\Ket{u_{-\infty}^P(\lambda^+)} - i \, \overline{r^{pp}} \Ket{u_{+\infty}^P(\lambda^+)}  
            -  \overline{r^{ps}} \Ket{u_{-\infty}^S(\lambda^+)} \\
    \Ket{u_0^S(\lambda^-)} & =  -\Ket{u_{+\infty}^S(\lambda^+)} - \overline{r^{ss}} \Ket{u_{-\infty}^S(\lambda^+)}  
            - i \, \overline{r^{sp}} \Ket{u_{+\infty}^P(\lambda^+)} 
\end{aligned}.
\end{equation}
For propagating $S$ and evanescent $P$ waves in the half-space, the unitary relations take a more complicated form \cite{kennett78}:
 $  \overline{r^{ss}} \, r^{sp} = i \, \overline{r^{sp}}  $.
With the aid of equation (\ref{irel}), this implies
 $ \overline{r^{ss}} \Ket{u_0^S(\lambda^+)} = - \Ket{u_0^S(\lambda^-)} . $
For $z<z'$, using the unitary relations, the resolvent discontinuity can  be put into the following form:
\begin{equation}
 \begin{split}
\mathbf{G}^+ -  \mathbf{G}^-  = & -\left[ i  \Ket{u_0^P(\lambda^+)} +  \Ket{u_0^P(\lambda^-)}\right] \Bra{u_{+\infty}^P(\lambda^+)}  \\
  &        + \Ket{u_0^S(\lambda^+)}\left[\Bra{u_{-\infty}^S(\lambda^+)} + \overline{r^{ss}} \Bra{u_{+\infty}^S(\lambda^+)} \right]
\end{split}.
\end{equation}
Using the symmetry relation $ r^{pp} - \overline{r^{pp}} = i \, \overline{r^{sp}} $,
 the following relation is established:
\begin{equation}
i \ket{u_0^P(\lambda^+)} + \ket{u_0^P(\lambda^-)} = -\overline{r^{sp}} \Ket{u_0^S(\lambda^+)},
\end{equation}
which in turn implies:
\begin{equation}
\mathbf{G}^+ -  \mathbf{G}^-  =  \frac{i}{\sqrt{\lambda^+}}  \Ket{u_0^S(\lambda^+)}  \Bra{u_0^S(\lambda^+)}  .
\end{equation}
This last equation has the desired form and shows that in the range of $\lambda$ considered there is only one
generalized eigenfunction. In the half-space, it consists of the sum of incident, reflected propagating $S$ wave
  and a reflected evanescent $P$ wave. Note that for some $\lambda$ in the range 
$\left]\beta_{\infty}^2 k^2, \alpha_{\infty}^2 k^2  \right[$, it may happen that the vector $\ket{u^P_{+\infty}}$
satisfies the zero traction condition at the free surface. Such a situation can be considered
as accidental, since a slight modification of the thickness or velocity in one the layers is likely
to make the mode disappear (Y. Colin de Verdi\`ere, personnal communication) . Such possible complications
will  be neglected.  \\
(3) For $\beta^2 k^2 > \lambda$, both $P$ and $S$ waves are evanescent in the half-space $z>0$.
 In this case the reflection matrix $R$ is simultaneously symmetric and hermitian, so that all its
 coefficients are real.  Further, using the  relations:
  \begin{equation}
\label{irel2}
\begin{aligned}
 \ket{u^{P}_{\pm\infty}(\lambda^-)}  = &  -i \ket{u^{P}_{\pm\infty}(\lambda^+)} &  \text{ and } 
   \ket{u^{S}_{\pm\infty}(\lambda^-)}  = & -i \ket{u^{S}_{\pm\infty}(\lambda^+)}
  \end{aligned}   ,
\end{equation} 
one establishes:
\begin{equation}
    \ket{u_0^{P,S}(\lambda^-)} =  -i  \ket{u_0^{P,S}(\lambda^+)}  ,
\end{equation}
which seems to implies that the resolvent discontinuity vanishes identically. 
This is not true as there can exist values of $\lambda$ for
which the reflection coefficients have poles. For these particular points of the spectrum, 
 the kets $\ket{u_0^{P,S}(\lambda^+)}$ can be written
solely as linear combinations of the kets  $\ket{u_{+\infty}^{P,S}(\lambda^+)}$ \cite{kennett83}. 
This means that there exist eigenvectors with
finite energy that satisfy the traction free condition, which are the well-known Rayleigh surface waves.
To each pole, we can associate
a one dimensional eigenspace  spanned by a normalized Rayleigh wave eigenvector.
For a given value of the horizontal wavenumber $k$, there can be $n$ distinct solutions denoted
 by $\lambda_0 < \lambda_1 < \cdots < \lambda_n$. The index serves to label the different surface waves
mode branches.  The index $0$ refers to the fundamental Rayleigh mode.
\subsection{Eigenfunction expansions}
We have now identified all the singularities of the resolvent on the real axis and are in position
to write down the completeness relation for 2-D in-plane problems  using formula (\ref{stone1}).
In the frequency domain, introducing the new variable $\omega^2=\lambda^+ $, one obtains:
\begin{equation}
\label{resolvom}
\begin{aligned}
\mathbf{I} = & \int\limits_{-\infty}^{+\infty} dk \sum_{n} \Ket{e^R(\omega_n)}\Bra{e^R(\omega_n)}  
             +  \int\limits_{-\infty}^{+\infty} dk \int\limits_{|\beta_{\infty} k|}^{|\alpha_{\infty} k|} d\omega \Ket{e^S(\omega)}\Bra{e^S(\omega)}   \\
            & + \int\limits_{-\infty}^{+\infty} dk \int\limits_{|\alpha_{\infty} k|}^{+ \infty} d\omega \left( \Ket{e^S(\omega)}\Bra{e^S(\omega)} + 
           \Ket{e^P(\omega)}\Bra{e^P(\omega)}  \right) ,
\end{aligned}
\end{equation}
where the compound generalized eigenvectors $\ket{e^{S,P,R}}$ are defined as:
\begin{equation}
\label{defeigenvec}
\begin{aligned}
   \braket{r | e^{S,P}(\omega)} = &  \sqrt{2}  \frac{e^{i k x}}{\sqrt{2 \pi}} \braket{z|u_0^{S,P}(\omega^2)}   
  \text{ and } \braket{r | e^{R}(\omega)} = &  \frac{e^{i k x}}{\sqrt{2 \pi}}  \braket{z | u^R(\omega_n)} \\
\end{aligned}.
\end{equation}
In the first of equations (\ref{defeigenvec}), the $\sqrt{2}$ factor is the result of introducing of the new variable
 $\omega$, whereas the last equation defines properly surface waves with unit-energy normalized  $z$-eigenfunction.  
Although this is implicit in equation (\ref{defeigenvec}), 
the reader should keep in mind that all eigenvectors are functions of the horizontal wavenumber $k$.
This  applies to the eigenfrequencies $\omega_n$ as well as the number of branches.  
As  in the case of the homogeneous space, it is possible to replace the frequency integral by a vertical
 wavenumber integral in equation (\ref{resolvom}) to obtain:
\begin{equation}
\label{resolvom2}
\begin{aligned}
\mathbf{I} = & \int\limits_{-\infty}^{+\infty} dp_x \sum_{n} \Ket{\psi^R(\omega_n)}\Bra{\psi^R(\omega_n)} 
            +  \int\limits_{-\infty}^{+\infty} dp_x \int\limits^0_{-\infty} dp_z \Ket{\psi^S(p_x,p_z)}\Bra{\psi^S(p_x,p_z)} \\
          & +    \int\limits_{-\infty}^{+\infty} dp_x \int\limits^0_{- \infty} dp_z  \Ket{\psi^P(p_x,p_z)}\Bra{\psi^P(p_x,p_z)} .
\end{aligned}
\end{equation}
 The eigenvectors $ \ket{\psi^{P,S,R}}$ are obtained by application of the transformation $U^{\dagger}$ to the kets
$ \ket{e^{S,P,R}}$. The eigenvectors  $\ket{\psi^{P,S}}$ correspond to properly normalized incoming $P$ and $S$ waves
exactly as given in equations (\ref{psips})-(\ref{hatps}) together with their reflections from the stack of layers. This is a general result in scattering
theory: the eigenvectors of the medium+scatterer can be obtained by calculating the complete response of the scatterer to the
unperturbed eigenvectors \cite{reed79,deverdiere06}.  This is basically what the double integrals of equation (\ref{resolvom2}) say. 
Note that by calculating  the scattering of plane waves,
one does not obtain directly the surface waves. They show up as poles of the reflection coefficient located on the
  real axis and,  in that  sense, can be compared to the bound states in scattering theory.  

\section{Application to diffuse fields} \label{application}
A diffuse field is defined as a random field where all the modes are equally represented. More precisely,
according to Ref.\onlinecite{weaver82}, it is a narrow-band signal which is a sum of normal modes of the system excited 
randomly at equal energy:
\begin{equation}
\label{modsum}
\begin{split}
   u(\mathbf{x},t) = & \sum\limits_{n,R,L}\iint\limits_{\omega_n \in B} \frac { dp_x dp_y}{2 \pi} A^{R,L}_n(p_x,p_y) \psi^{R,L}_n(z) 
  e^{i (p_x x + p_y y)} e^{- i\omega_n^{R,L}(p_x,p_y) t} \\
  &  + \int\limits_{p_z<0} dp_z \sum\limits_{P,S}  \iint\limits_{\omega \in B} \frac { dp_x dp_y}{2 \pi}  A^{P,S}(p_x,p_y,p_z) 
   \psi^{P,S}_{p_z}(z)   e^{i (p_x x + p_y y)}   e^{-i\omega^{P,S}(p_x,p_y,p_z) t}.
\end{split}
  \end{equation}
In equation (\ref{modsum}), $B$ denotes the  frequency band of the signal and $u$ refers to the complex analytic
signal whose real part is the measured displacement vector.
In a diffuse field, the amplitudes $A_n^{R,L}$ and $A^{P,S}$ are assumed to be  white-noise random processes:
\begin{equation}
 \label{an}
 \begin{split}
   \langle A^{R,L}_n A^{P,S} \rangle = & 0 ,\\  
 \langle A^{R,L}_n (p_x,p_y)   A^{R,L}_m (p'_x,p'_y)^{\star} \rangle = &   \sigma^2 \delta^{R,L} \delta_{nm}  \delta(p_x -p'_x) \delta(p_y -p'_y), \\
  \langle A^{P,S} (p_x,p_y,p_z) A^{P,S}(p'_x,p'_y,p'_z)^* \rangle  = & \sigma^2 \delta^{PS} \delta(p_z -p'_z)  
   \delta(p_x -p'_x) \delta(p_y -p'_y)  .
\end{split}
\end{equation}
In the case of a semi-infinite medium, the sum over all modes involves a continuous index
that labels the eigenfunctions of the continuous spectrum and a discrete index that refers to the
surface wave mode branch.

Formulas (\ref{modsum})-(\ref{an}) are now applied to the calculation
of  the vertical to horizontal kinetic energy ratio of a diffuse field in a stratified
half-space. The vertical (or horizontal) kinetic energy density at central frequency
$\omega_0$ is obtained by means of the Wigner distribution of the complex analytic wavefield:
\begin{equation}
  \label{eq:wigner}
  E_z(t,\tau, \mathbf{x}) = \frac12 \rho(z)  \left\langle \partial_t u_z(t+\tau/2, \mathbf{x}) 
    \partial_t u_z(t-\tau/2, \mathbf{x})^* \right\rangle.
\end{equation}
In equation (\ref{eq:wigner}), the brackets denote an ensemble average. Inserting the spectral
decomposition (\ref{modsum}), applying the equipartition principle (\ref{an}),
  and taking a Fourier transform with respect to the time
 variable $\tau$, the local vertical kinetic energy  density of a diffuse field at circular frequency $\omega_0$
is obtained: 
 \begin{equation}
\label{ez}
\begin{split}
 E_z(\omega_0,z) = & \frac{\rho(z)\omega_0^2 \sigma^2}{8\pi^2}   
    \sum\limits_{n,R,L}\iint\limits_{\mathbb{R}^2}  dp_x dp_y \left| \psi^{R,L}_{n,z}(z) \right|^2  
     \delta \left(\omega_0 - \omega^{R,L}_n(p_x,p_y) \right) \\
     & +  \frac{\rho(z)\omega_0^2 \sigma^2}{8\pi^2} \int\limits_{p_z<0} dp_z \sum\limits_{P,S} 
  \iint\limits_{\mathbb{R}^2}  dp_x dp_y | \psi^{P,S}_{p_z,z}(z)|^2  
     \delta \left(\omega_0 - \omega^{P,S}(p_x,p_y,p_z) \right).
\end{split}
\end{equation}
In equation (\ref{ez}), the delta functions represent the  density
of states of the surface and body waves, which are closely related to the imaginary part of the
Green's function \cite{sheng95}. The link between the Wigner distribution of the wavefield and the Green's function 
is the theoretical basis of the Green's function reconstruction in diffuse fields \cite{tiggelen03,weaver01b,deverdiere06}.
Introducing cylindrical and spherical coordinates in the double and triple integrals of equation (\ref{ez}), respectively,
one finds:
 \begin{equation}
\label{ez2}
\begin{split}
 E_z(\omega_0,z) = &  \sum\limits_{n,R,L} \frac{\rho(z)\omega_0^3 \sigma^2}{4\pi c^{R,L}_n u^{R,L}_n} \left| \psi^{R,L}_{n,z}(z) \right|^2 \\  
     & + \sum\limits_{P,S} \frac{\rho(z)\omega_0^4 \sigma^2}{4\pi v_{p,s}^3} \int\limits_{\pi/2}^{\pi} d\theta  \sin{\theta}  
   | \psi^{P,S}_{p_z,z}(z)|^2\bigg\vert _{\displaystyle p_z=\frac{\omega_0}{ v_{p,s}} \cos \theta, }   .
\end{split}
\end{equation}
In equation (\ref{ez2}), we have introduced the following notations: $c_n$ and $u_n$ are the phase and group
velocities of the n$^{th}$  surface wave mode, respectively; $v_{p}$-($v_s$) stands for 
$\alpha_{\infty}$-($\beta_{\infty}$).
The calculation of the eigenfunctions and the remaining
sum and integral have to be performed numerically. 
 
\begin{figure}
   \includegraphics[width=0.8\linewidth]{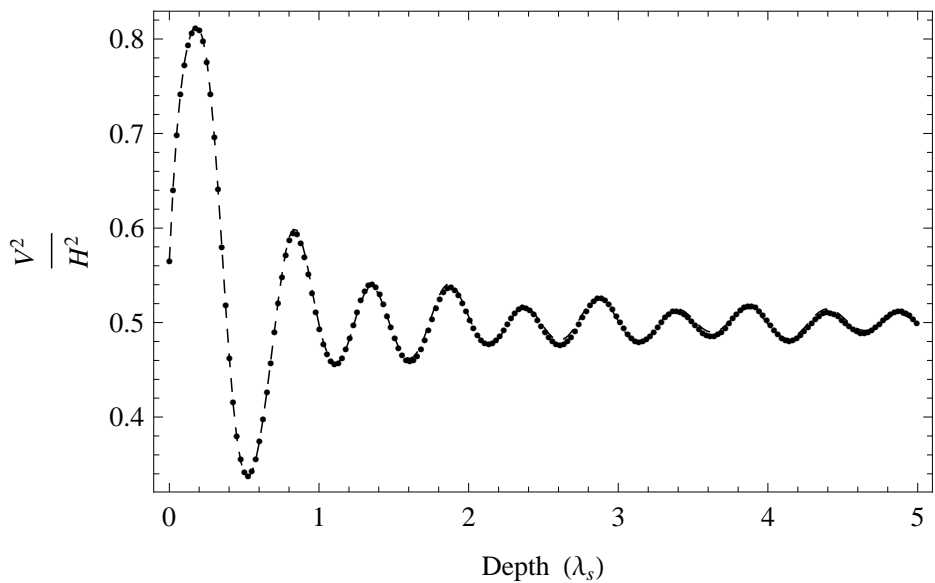}
  \caption{Depth dependence of the vertical to horizontal kinetic energy ratio  near the free surface of a Poisson half-space.  Dots: locked-mode approximation. Dashed line: generalized eigenfunctions summation. The depth unit is the shear wavelength $\lambda_S$ }
\label{equihalf}
\end{figure}

In Figure (\ref{equihalf}), we consider the vicinity of a free surface which has been
previously investigated by several authors \cite{weaver85,hennino01,tregoures02}.
The calculations were  performed both with formula (\ref{ez2}) and  with a locked-mode approximation where the medium is bounded
at great depth by a rigid boundary where the displacements vanish exactly.
In the latter case, the eigenvalue problem in the depth coordinate only has a discrete spectrum, which is standard.
For more details on the locked-mode technique and geophysical applications, the reader is referred to Ref.\onlinecite{nolet89}.
 The outcome of the two calculations for a 3-D medium are superposed in Figure \ref{equihalf}. In the locked-mode
technique, the lower boundary is at a depth of 16 shear wavelengths.  A total of 32 Love and 50 Rayleigh modes
were found. The agreement is very satisfactory
and confirms the validity of our approach. Our work shows that the generalized eigenfunctions of the continuum can be treated
like standard normal modes, albeit with a continuous index.
 The results presented in Figure (\ref{equihalf})  were first obtained in Refs. 
\onlinecite{hennino01},\onlinecite{tregoures02}, using the eigenfunctions of a thick plate, the so-called Lamb modes. 
Our calculations illustrate the fact that
the elastodynamic operator is in the limit point case at $+\infty$ according to the classification of 
Ref.\onlinecite{krall89}.
This simply means that independent  of the self-adjoint  boundary condition imposed at the lower boundary 
-Neumann, Dirichlet or mixed-, the
eigenfunctions converge to a common limit as the depth of that boundary tends to $\infty$. A few wavelengths away from
the free surface, the kinetic energy ratio oscillates around 0.5, which is the expected value
in a homogeneous half-space.

\begin{figure}
  \includegraphics[width=0.8\linewidth]{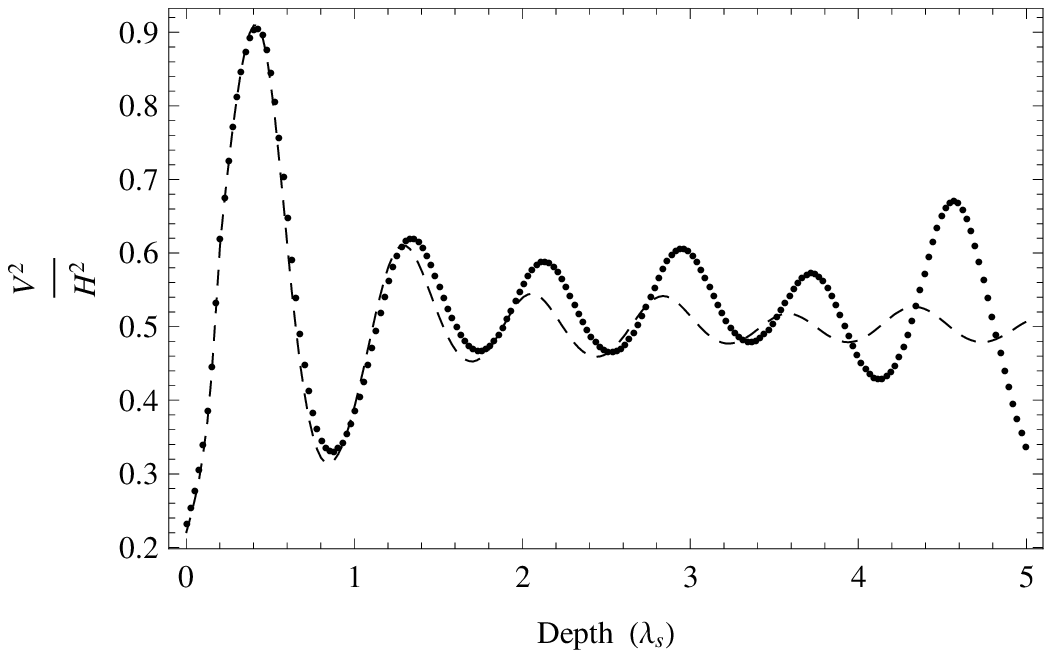}
   \includegraphics[width=0.8\linewidth]{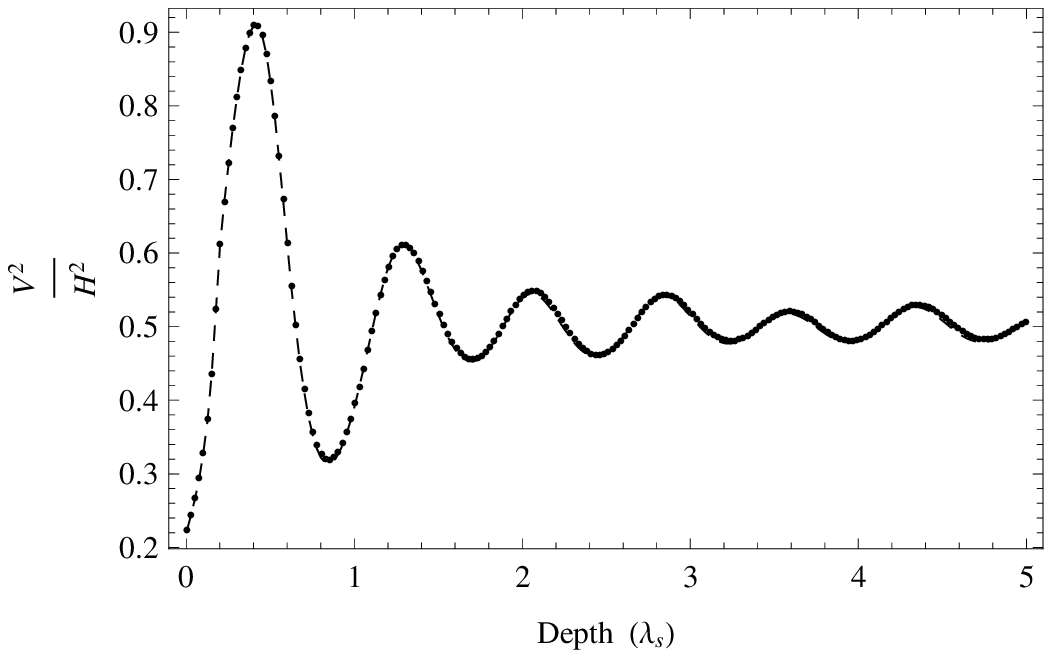}
  \caption{Depth dependence of the vertical to horizontal kinetic energy ratio in the presence of a soft layer.
 Dots: locked-mode approximation. Dashed line: generalized eigenfunctions summation. The depth unit is the shear
 wavelength $\lambda_S$.
 In the locked mode approximation, the lower boundary is located at a depth of 5.25$\lambda_s$ (Top)  and 42$\lambda_s$ (Bottom),
 respectively.}
 \label{layer}
\end{figure}

We now investigate the case of a soft layer overlying a homogeneous space.
In the layer, the $P$ and $S$ wave velocity is one third smaller than in the half-space and
the density is reduced by a factor 2. The vertical to horizontal kinetic energy 
ratio is plotted as a function of depth in Figure \ref{layer}. The unit depth is the shear wavelength inside
the layer and the interface between the layer and the half-space lies at about 0.2 unit depth.
In the locked-mode approximation,
  two situations were considered where the rigid boundary is located at 5.25 and 42  wavelengths below the
bottom of the layer. In the former case, 7 Love and 11 Rayleigh modes were found, while in the latter
case we identified a total 56 Love and 88 Rayleigh modes. In the generalized eigenfunction expansion
method, only the fundamental Love and Rayleigh modes are present. Thus, in the
locked-mode method, the higher Love and Rayleigh modes serve as an approximation for
the propagating $P$ and $S$ waves incident from below the layer. 
As illustrated in Figure \ref{layer}, the agreement between the two methods is extremely good and
as expected, improves as the depth of the rigid boundary increases. 
In the vicinity of  the layer,  the vertical to horizontal  kinetic energy ratio shows 
rapid variations and, at greater depth, converges towards the ratio of a homogeneous
space. 

\begin{figure}
      \hspace*{0.57cm} \includegraphics[width=0.75\linewidth]{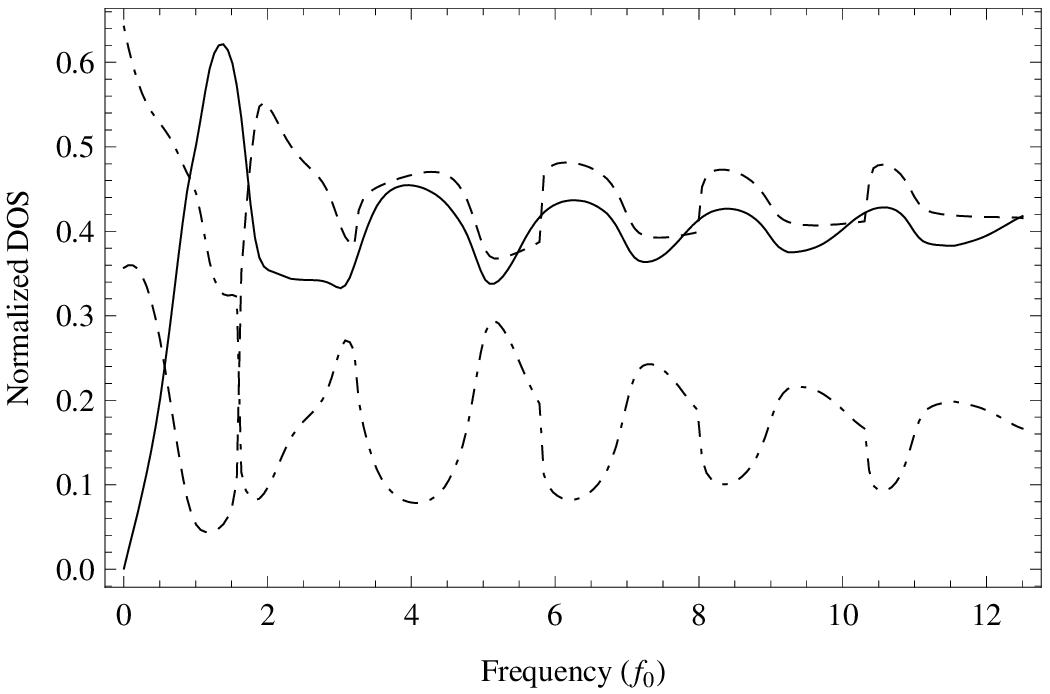}
     \includegraphics[width=0.8\linewidth]{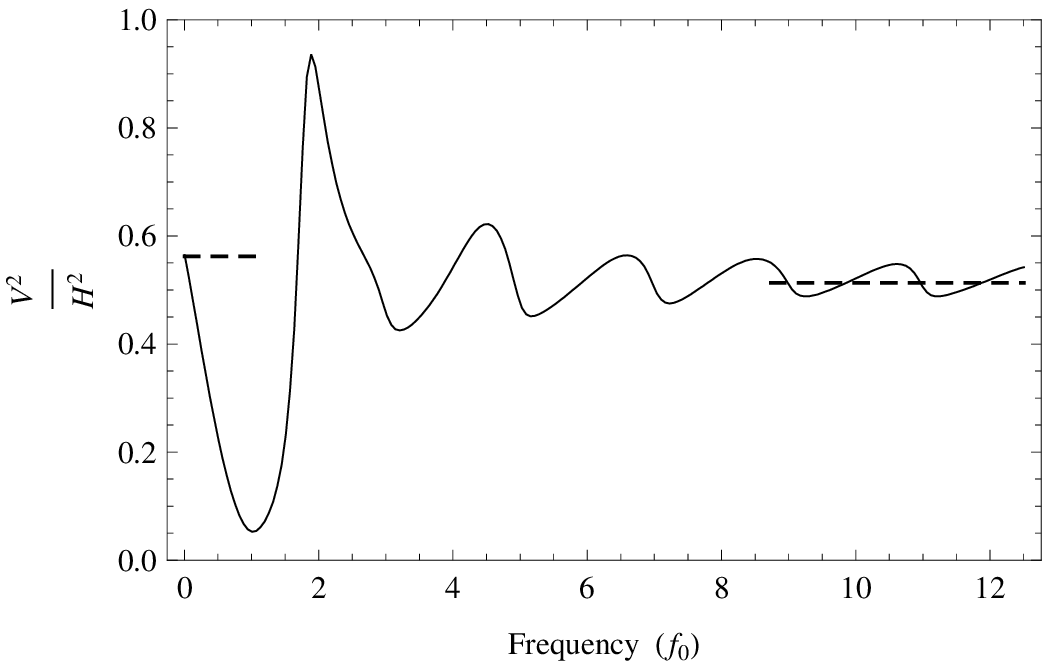}
  \caption{Properties of a diffuse field at the surface of a half-space with a superficial low-velocity, low-density layer.
 Top: Normalized density of states of Rayleigh (dashed line), Love (solid line) and bulk waves (dash-dotted line)
 as a function of normalized frequency. 
The  resonance frequency $f_0$ of vertically propagating $S$ waves in the layer is the unit frequency.
Bottom: Frequency dependence of the vertical to horizontal kinetic energy ratio.
 Dashed lines: low- and high-frequency asymptotics. 
Solid line: generalized eigenfunctions summation. }  \label{softlayer}
\end{figure}

In Figure \ref{softlayer}, we investigate the frequency dependence of the $V^2/H^2$ kinetic energy ratio
at the surface of a solid with a soft layer at the surface, where
the shear velocity and density  are reduced by a factor 2 with respect to the underlying
Poisson half-space. The ratio of longitudinal to shear wavespeeds in the layer is taken to be 2.5.
 To facilitate the  interpretation
of the  $V^2/H^2$ calculations, we show in the Top panel the normalized density of states of body, Rayleigh and Love
waves as a function of frequency. The frequency unit is the fundamental resonance 
frequency of the layer for vertically propagating shear waves, $f_0 = \beta/4h$, where $\beta$ is the  shear wavespeed 
and $h$ is the layer thickness.
In Figure \ref{softlayer}, we have also indicated high- and low-frequency asymptotics of the $V^2/H^2$ energy ratio.
At high frequency, we expect the waves to be largely insensitive to the properties of the underlying half-space.
We  therefore calculate an approximate high-frequency $V^2/H^2$ ratio by replacing the true model with 
a  simple half-space where the $P$ to $S$ velocity ratio equals  2.5. The expected value is 0.513, in good agreement
with the full calculations. 
  At low-frequency, we can neglect the presence of the shallow layer to recover the classical 0.56 ratio
at the surface of a Poisson solid. The  $V^2/H^2$ ratio  is slightly smaller at high frequencies  mainly  because 
of  the increased amount of horizontally polarized shear waves. To the contrary, the $V^2/H^2$ energy ratio
of the Rayleigh wave tends to increase with the $P$ to $S$ velocity ratio. Close to the resonance frequency $f_0$ of the layer, the
vertical to horizontal kinetic energy ratio drops dramatically. Careful analysis reveals that this is caused 
by the  increasing contribution  of the fundamental Love mode to the density of states at the surface.
Around twice the resonance frequency, the $V^2/H^2$ ratio presents a marked overshoot which is due to the appearance
of Rayleigh waves higher modes (i.e. body waves trapped in the low velocity layer) which are preferentially polarized
on the vertical axis. At high frequencies, the energy density is largely dominated by the Rayleigh and Love waves
trapped in the low-velocity layer. The fluctuations of the local density of states slowly decrease with
increasing frequency. 
\section{Conclusion} \label{conclusion}
In this work, we have shown that the definition of a diffuse field as a white noise in modal space
can be applied  equally well to closed  \cite{weaver82} and  open systems \cite{weaver04}.
In the framework of the  spectral theory presented in this paper, discrete,
continuous or mixed spectra can be treated on the same footing. 
Other applications of the generalized normal mode expansion
could be envisaged, such as the  calculation of synthetic seismograms
for geophysical applications \cite{maupin96}. The theory could
also be used to give more rigorous foundations to empirical
civil engineering techniques.
In particular, the large drop of the vertical
to horizontal kinetic energy ratio in diffuse fields close to the resonance
frequency of a low-velocity layer sounds reminiscent of the so-called
Nakamura's method used for site effect evaluation with ambient noise \cite{bard99}.
Although extremely popular, the limitations of the technique are still to be understood.
 The diffuse field concept  offers a potentially useful tool  for this purpose.
\begin{acknowledgments}
I would like to express my gratitude to Pr. Y. Colin de Verdi\`ere for his lectures on mathematical 
topics closely related to the present article. This paper has also been influenced
by discussions with Dr. H. Nakahara, who suggested to examine the wave content
of diffuse fields. 
\end{acknowledgments}
\appendix
\section{\label{dirac} Summary of Dirac formalism}
In this appendix, we summarize the bra-ket notations used in this paper.
A vector or ket  of an abstract vector space  will be denoted by:
 $\Ket{e} $.
The representation of the vector in position space is written as:
$
  \label{braxrep}
  \braket{x | e} = e(x).
$
To each ket of our space of function, we associate a bra denoted by
$
 \Bra{e} = \Ket{e}^{\dagger} .
$
In position space, a   bra has representation
 $
  \braket{e | x} = e(x)^*,
$
where $^*$ denotes complex conjugation. Mathematically speaking, a bra would be more appropriately
defined as a linear functional acting on a space of test functions. However, we will not insist on these
technicalities. We shall also make use of the conjugated versions of kets such that:
$   \braket{x | e^*} = e(x)^* .$
The scalar product between two vectors $\ket{e}$ and $\ket{f}$ is denoted by
\begin{equation}
  \label{scalprod}
   \braket{e | f} = \int dx \rho(x) e_i(x)^*  f_i(x) ,
\end{equation}
where a summation over the repeated index $i$ is implied.
The completeness relation or resolution of the identity  for a set of eigenvectors $\Ket{e_p}$ is written as:
\begin{equation}
 \label{completeness}
  \sum_{p} \Ket{e^{ }_p} \Bra{e^{ }_p}  = \mathbf{I},
\end{equation}
where $\mathbf{I}$ denotes the identity in the abstract vector space, and $p$ denotes
a label running over the whole set of eigenfunctions.
 Equation (\ref{completeness}) introduces the outer  product between a bra and a ket.
 The outer product of two (properly normalized) eigenvectors is an orthogonal projector on the
 subspace generated by $\Ket{ e} $.  
In the position representation, equation (\ref{completeness}) reads 
\begin{equation}
   \sum_{p}   \braket{x \Ket{ e_p^{ }  } \Bra{  e_p^{ } } x'}  = 
       \delta_{ij} \frac{\delta(x - x')}{\rho(x')}. 
\end{equation} 
The appearance of the weighting function $\rho$ in the denominator is consistent with the 
definition of the scalar product (\ref{scalprod}).
The matrix elements in position and polarization space of a general outer product between a ket 
$\Ket{e}$ and a bra $\Bra{f}$ are given by
\begin{equation}
    \braket{x \Ket{ e_{ }^{ }} \Bra{ f_{\ }^{ }}  x'} =
       e_i(x)  f_j(x')^*   .        
\end{equation} 
The matrix elements of an abstract  operator $\mathbf{L}$  in position space are given by the
 kernel of  an integral operator. If $\mathbf{L}$ represents a differential operator, it will be assumed
 to be diagonal in position space.

\end{document}